\documentclass{article} 
\usepackage{nips12submit_e,times}

\usepackage[pdftex]{graphicx}
\usepackage{url}
\usepackage{amssymb}
\usepackage{multirow}
\usepackage{subfigure}

\newcommand {\comment}[1] {}
\newcommand {\comments}[1] {}

\newcommand{\ie}{\textit{i.e.}~\/}
\newcommand{\eg}{\textit{e.g.}~\/}

\newlength \figwidth
\if@twocolumn
\else
\fi

\title{Toward improving the visual stimulus meaning for increasing the P300 detection}

\author{Hubert Cecotti\\
Northern Ireland Functional Brain Mapping Facility\\
Ulster University, Magee Campus, Derry $\sim$ Londonderry, UK
\And
Bertrand Rivet\\
GIPSA-lab CNRS UMR 5216, Grenoble Universities\\
F-38402 Saint Martin d'H\'eres, France
}

\nipsfinalcopy 

\begin{document}

\maketitle

\begin{abstract}
The P300 speller is a well known Brain-Computer Interface paradigm that has been used for over two decades. A new P300 speller paradigm (XP300) is proposed. It includes several characteristics: \textit{(i)} the items are not intensified by using rows and columns, \textit{(ii)} the order of the visual stimuli is pseudo-random, \textit{(iii)} a visual feedback is added on each item to increase the stimulus meaning, which is the main novelty. XP300 has been tested on ten healthy subjects on copy spelling mode, with only eight sensors. It has been compared with the classical P300 paradigm (CP300). With five repetitions, the average recognition rate across subjects is 85.25\% for XP300 and 77.25\% for CP300. Single-trial detection is significantly higher with XP300 by comparing the AUC (Area Under Curve) of the ROC (Receiver Operating Characteristic) curve. The mean AUC is 0.86 for XP300, 0.80 for CP300. More importantly, XP300 has also been judged as more convenient and user-friendly than CP300, hence being able to allow longer sessions.
\end{abstract}

\section{Introduction}

The performance of a a Brain-Computer Interface (BCI) is related both to the performance of the signal processing techniques that are used for assigning some  electroencephalogram (EEG) signal to a command and to the possibility of the user to adapt him/herself to the system over time. Indeed, advanced machine learning, signal processing and classifier techniques have been widely used for improving BCIs, \textit{e.g.}~\cite{mull08}. In spite of these recent improvements in the BCI community, several obstacles remain to fully transfer laboratory-demonstrated BCIs to real commercial/clinical applications. Alternative systems like eyetrackers or BCI based on Steady-State Visual Evoked Potentials could be effective solutions for spelling applications. These systems are even more relevant as the efficiency of the P300 speller depends on eye gaze~\cite{brun10}. In regard to these recent studies, the BCI should be chosen in relation to the gaze control capability of the patient. However, the best judge shall remain the user/patient with the BCI part integrated in a complete functional system. Therefore, the improvement of the P300 speller shall still be an active research topic. We consider in this work the BCI P300-speller. This paradigm allows people to select symbols where each symbol is depicted in a cell of a matrix on a computer screen. It is based on the oddball paradigm to generate event-related potentials (ERPs), like the P300, on targets selected by the user. The classical P300-Speller consists of a $6 \times 6$ matrix, which the different items that can be selected by the user~\cite{far88}. During the experiments, the user has to focus on the character she/he wants to spell. When the user focuses on a cell of the matrix, \ie the item the person wants to select, it is possible to detect a P300, which corresponds to a positive deflection in voltage at a latency of about 300 ms relative to the stimuli onset in the EEG signal. This deflection is time-locked to the onset of the cell intensification. The rows and columns are intensified randomly to generate ERPs. In the original implementation of the speller, row/column intensifications are block randomized in 12 events (6 rows and 6 columns). This set of 12 intensifications is repeated $N_{epoch}$ times for each character. Hence, $2N_{epoch}$ possible P300 responses should be detected for the recognition of one character.

The main challenge for the P300 paradigm is to provide a reliable and fast performance in a domestic or clinical environment, \ie outside of the laboratory. To achieve such goals, the sensors can be ideally chosen to reduce the preparation time, the cost of an amplifier and the EEG cap~\cite{ceco10c}. Most of the available systems lacks robustness over time and across subjects: the performance does not meet user's requirements due to unadapted end-user interface. However, there exist few exceptions as in~\cite{vaug06}, where a late stage ALS (Amyotrophic lateral sclerosis) patient could use a P300-Speller at home developed by the Wolpaw lab. The intendiX solution has been proposed by g.tec in 2009. This BCI is designed to be installed and operated by caregivers or the patient's family at home. According to the g.tec company, the performance for the majority of healthy users during their first trial is estimated to a spelling rate of 5 to 10 characters per minute (cpm). This BCI is proposed as a whole package (software, amplifier, caps,...), which provides a global solution~\cite{gtec}.

Most of the improvements in the P300-Speller have been achieved at the signal processing like spatial filtering~\cite{Blankertz2008,rivet09} and/or detection level with techniques like Support Vector Machines~\cite{rako08}, neural networks~\cite{ceco10a} or Bayesian linear discriminant analysis~\cite{ceco10c,hoff08}.
In contrast, the P300-Speller graphical user interface (GUI), in its matrix form, has not evolved much for more than two decades in spite of some studies, \eg~\cite{all03,sel06}. Some improvements have been proposed as in~\cite{taka09}, where the color of the flickering matrices should be green/blue. To extend the user-centered approach known for signal detection, many problems should be addressed for the design of GUI. Indeed, a poorly designed GUI can be an obstacle for an ideal detection of the P300.

In the present study, we show some efficient and easy solutions for solving problems demonstrated in the literature about the P300 speller paradigm. In addition, we propose the addition of a visual feedback for each visual stimulus on the possible targets. The paper is organized as follows. The main novelties of the P300 speller, the GUI and the signal processing methods, are presented in the second and third section, respectively. The experimental protocol is given in the fourth section. Results are detailed and then discussed in the last sections.

\section{P300-Speller paradigm}

\begin{figure}
\centering
\includegraphics[width=0.9\columnwidth]{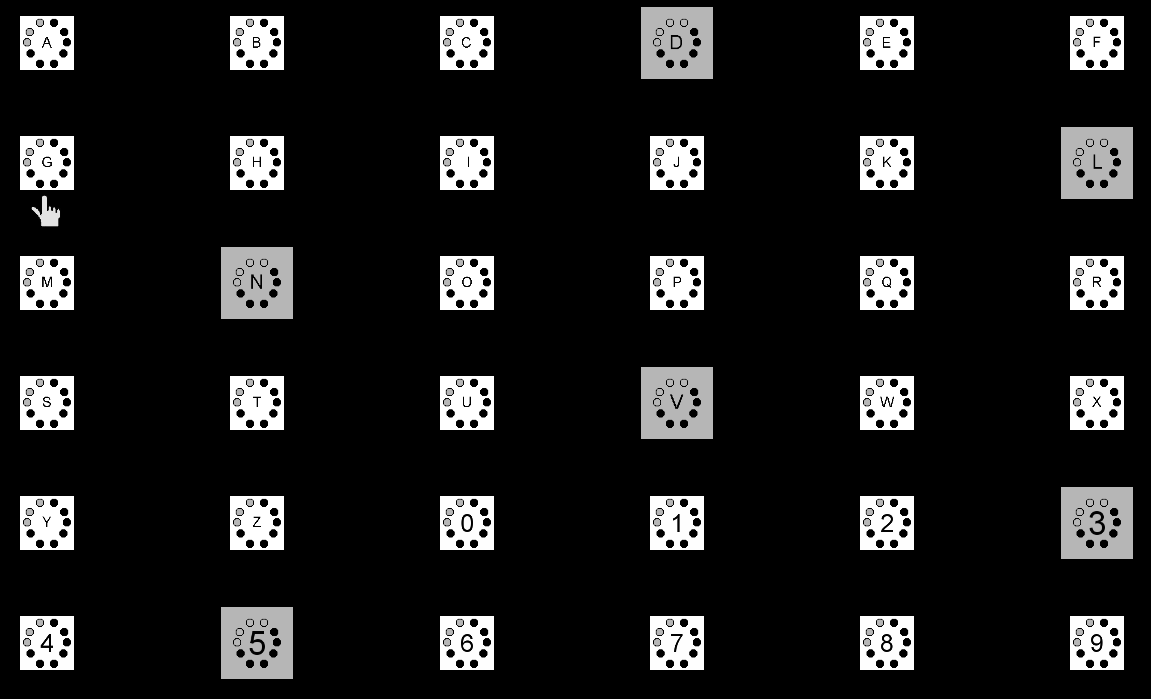}
\caption{GUI overview of the proposed XP300.
In this example, $\{$D, L, N, V, 3, 5$\}$ are intensified at the 4th repetition. With proposed XP300 paradigm, \textit{(i)} the intensified items do not correspond to a row or a column of the matrix and \textit{(ii)} the presence of the feedback (small bullets) for counting visual stimuli.}
\label{fig:gui}
\end{figure}

Two paradigms were evaluated: the classical P300 speller (CP300) and the proposed P300 speller paradigm (XP300), which is depicted in Fig.~\ref{fig:gui}. These two paradigms differ only in the following three parameters: the place of the flashes (section~\ref{sec:vs1}), the order of the flashes (section~\ref{sec:vs2}), and the visual feedback for counting flashes (section~\ref{sec:vs3}).

\subsection{Visual stimuli location}
\label{sec:vs1}

In the P300 paradigm, only flashes of groups containing the attended item should elicit a P300. In the original implementation of the P300-BCI, and in its most recent implementations, flashed items are grouped as rows and columns (RC). One major problem with the RC paradigm is the classical presentations of the different flashes as they involve frequent error around the target. Furthermore, the P300-BCI GUI has not really evolved for 22 years~\cite{far88}. Although the size, the color and other parameters of the interface have been investigated, the RC paradigm is still one major invariant in the literature. An alternative to the RC paradigm, the checkerboard paradigm (CB), has been proposed~\cite{town10}. It is a combination of two checkerboards to avoid the well known confusion problems in the neighborhood of the target~\cite{fazel2007}. Both the online accuracy and the information transfer rate (ITR) were also significantly higher for the CB than for the RC. They have proved that the RC paradigm has a direct impact on the performance. Thus, improving the flashing paradigm will improve the performance. However, in~\cite{tan10}, a comparison across nine healthy subjects of four stimulus presentation modes revealed that only complex highlighting effects composed of brightness enhancement, rotation, enlargement and a trichromatic grid overlay in combination with row-column spatial arrangements of simultaneously highlighted objects improved single sub-trial classification performance.

The RC paradigm is shown to be just a special case of a more generic model. In this model, we propose alternatives to the RC paradigm, which offer relevant properties for P300-BCIs. We denote by $N \times N$ the size of the matrix $M$ that contains the different symbols. $M(i,j)$ represents the symbol at the $i^{th}$ row and the $j^{th}$ column. In the RC paradigm, a target can be identified as the intersection of a row and a column after $2N$ flashes, where $N$ flashes occur on the matrix rows, and the other $N$ flashes are on the matrix columns.
Let  $R$ and $C$ denote the two $N\times N$ matrices containing the items to be intensified at the same time.
For instance with $N=6$,\\

We denote by $R$ and $C$, the matrices of size $N \times N$ that contain the indices of the different possible flashes. For instance, $\forall (a,b) \in \{1..N\}$, if $R(a,b)=f$ and the flash number $f$ must be flashed, then each cell $M(a,b)$ will be intensified. For $N=6$, we obtain:

{\small\begin{tabular}{c@{}c@{}}
$R$ = \begin{tabular}{cccccc}
1&1&1&1&1&1\\
2&2&2&2&2&2\\
3&3&3&3&3&3\\
4&4&4&4&4&4\\
5&5&5&5&5&5\\
6&6&6&6&6&6\\
\end{tabular}
& $C$ =
\begin{tabular}{cccccc}
1&2&3&4&5&6\\
1&2&3&4&5&6\\
1&2&3&4&5&6\\
1&2&3&4&5&6\\
1&2&3&4&5&6\\
1&2&3&4&5&6\\
\end{tabular}
\end{tabular}}
\normalsize

\noindent mean that all the cells $M(a,b)$ such that $R(a,b)=f$ (resp. $C(a,b)=f$) have to be intensified at the $f^{th}$ flash of the rows (resp. column). Each symbol of the screen $M$ can then be identified by two flashes extracted from $R$ and $C$: $M(R(a,b),C(a,b))$. Indeed, $\forall (a,b) \in \{1..N\}^2$, the couple $(R(a,b),C(a,b))$ is unique: it guarantees that with two flashes, a target can be selected on the matrix.
We propose now to create new matrices $\hat{R}$ and $\hat{C}$ in relation to a random distribution of the indices representing the symbols on the matrix $M$. It is possible to get a random distribution of the indices of the symbols represented on the screen while keeping the position of the items on the screen. Let $\hat{\mathbf{v}}$ be a vector obtained from a random permutation of $\{1,\cdots,N^2\}$. If $\hat{\mathbf{v}}=[1,\cdots,N^2]^T$, then $\hat{R}=R$ and $\hat{C}=C$, \textit{i.e.} it corresponds to the classical row/column intensification. With the newly created $\hat{R}$ and $\hat{C}$, the flashes are not necessary homogeneously represented as rows and columns, but they correspond to subsets of items on the screen.  Thus, the classical RC paradigm is not the only method to achieve an identification of a cell in the matrix with $2N$ flashes. This new method allows determining new flashing strategies in a random fashion. The surprising effect comes from both the random order of the flashes and the spatial locations of the flashed items on the screen. It is possible to go one step further by adding some constraints to the subsets of symbols that are simultaneously flashed. As indicated previously, one source of errors in the P300 speller can be the simultaneous flash of contiguous symbols. It is naturally the case in the RC paradigm where aligned symbols are flashed together. A constraint on the matrix $\hat{R}$ is proposed. It does not allow the flashes on 2 contiguous symbols aligned vertically and horizontally. A simple solution to this problem is to consider a row with each flash number (1 to N) and then  shift this row by one element on the second row. This process can be applied recursively for the other rows. Then $\hat{C}$ is obtained by considering the unicity of a couple $(\hat{R}(a,b),\hat{C}(a,b))$ ($\hat{C}$ is not unique). For $N=6$, one can obtain:

\small{\begin{tabular}{c@{}c@{}}
$\hat{R}$ = \begin{tabular}{cccccc}
1&2&3&4&5&6\\
6&1&2&3&4&5\\
5&6&1&2&3&4\\
4&5&6&1&2&3\\
3&4&5&6&1&2\\
2&3&4&5&6&1\\
\end{tabular}
& $\hat{C}$ =
\begin{tabular}{cccccc}
1&2&3&4&5&6\\
5&6&1&2&3&4\\
3&4&5&6&1&2\\
1&2&3&4&5&6\\
5&6&1&2&3&4\\
3&4&5&6&1&2\\
\end{tabular}
\end{tabular}}

\normalsize

The first row of $\hat{C}$ is set with a random permutation of a vector $u=(1,\ldots,N)$. Each value $S(\hat{R}(1,j),\hat{C}(1,j))$ is set to 1, with $1\leq j\leq N$.  

\subsection{Visual stimuli order}
\label{sec:vs2}

In the classical P300 paradigm, the flashes are randomly intensified by block of 12 flashes (6 rows + 6 columns). As the target is going to be intensified two times during a block of 12 flashes, two consecutive flashes can occur on the target. In such a situation, a user may fail to detect a second salient target occurring in succession if it is presented between 200-500ms after the first one, it is the attentional blink effect~\cite{raymond1992}. The expected P300 of the second flash can have a low amplitude, which can impair its detection. Indeed, the target epochs with a target to target interval of about 200ms are characterized by a severely reduced classification accuracy approaching thus the chance level~\cite{martens2009}. To optimize the stream of visual stimuli, we propose to separate the block of flashes: first the row, then the column. In addition, we consider a pause between the block of rows and the block of columns, and between each repetition, to avoid two consecutive flashes in a too short time. These pauses leads to a slower speller: 14 Inter Stimuli Interval  (ISI) instead of 12 for each repetition. During the experiments, the duration of the pause is equal to one ISI, \ie the duration between two visual stimuli is at least 2 ISI. It is then possible to show randomly the visual stimuli of $\hat{R}$ and then the visual stimuli of $\hat{C}$, instead of presenting randomly the visual stimuli corresponding to both $R$ and $C$. With $\hat{R}$ and $\hat{C}$, the notion of rows and columns disappears to the profit of a new notion: random sets of symbols. Hence, if we consider an ISI of 133ms, $N=6$ and if the flashes in $\hat{R}$ are set first randomly, and then those in $\hat{C}$ are set randomly, there will be in average more than 0.9s between two flashes on the target. 

\subsection{A feedback at each visual stimulus}
\label{sec:vs3}

The visual stimulus on each cell of the matrix has the following properties: the size of the cell is bigger ($\times 1.33$) than with non-intensified cells, the color of the cell becomes green, and a green dot is placed in the little holes around the target. Although alternatives like imaginary or real movement have been recently proposed as an alternative to mental counting~\cite{salvaris2010}, we propose a visual feedback for evaluating and counting the different flashes occurring for the selection of a character. It can also give a better estimation to the subject of when a character is going to be spelt. Indeed, the user shall have a cognitive activity, like counting, when a visual stimulus appears on the target. During long sessions, it can be quite difficult for the user to follow and count the different flashes. It is particularly the case for patient suffering from severe disabilities coupled with depression~\cite{bla87}. The depression can involve a lack of concentration and therefore a low amplitude of the P300 wave. Usually, to spell one character with a P300-BCI, the user needs at least 15 seconds, if we consider ten repetitions with an ISI of 133ms and 12 flashes. During this time the subject can lose his concentration and miss some flashes. When the user does not succeed in counting the expected number of flashes, she/he may become frustrated and this may change the subject's mood: if the number of flashes is not reached, the user can think she/he failed. The purpose of adding new events in the P300 paradigm is to increase the stimulus meaning~\cite{joh86}, to avoid frustration and to keep the subject more active during the experiment.

\section{Experimental protocol}

\subsection{Materials and subjects}
\label{sec:mat}

The EEG signal was recorded on ten volunteer healthy subjects. Subjects were wearing an EEG cap with 10 electrodes. $F7$ and $F8$ were dedicated to the ground and the reference, respectively. The signal was recorded on $O_1$, $O_2$, $P_3$, $P_4$, $P_7$, $P_8$, $P_Z$ and $FC_Z$. The amplifier was a FirstAmp (Brain Products GmbH) with a sampling frequency of 2kHz. A phototransistor was placed on the right corner of the screen for tagging directly in a precise way the events occurring on the screen with the recorded EEG signal (a flash synchronized with the flashes of the P300 speller was under the phototransistor). This method assured a precise synchronization between the EEG and the events on the screen. 

\subsection{Protocol}

The experiment was divided into four sessions: two for each paradigm CP300 and XP300. The order of the session was randomized for all subjects to avoid effects due to the fatigue or user training of the interface. For each session, each subject had to spell 20 characters on copy spelling mode. Sessions were set as a Wizard of Oz experiment: it was not specified if sessions were for calibration or test, characters were supposed to be written correctly. The number of repetitions was 10. The ISI was set to 133ms (66.66ms for the visual stimulus intensification duration), it corresponds to a frequency of 7.5Hz, \ie 8 frames on a screen with a VRR of 60Hz. At the end of the experiment, the subjects had to fill out a questionnaire and express some comments. The questionnaire was about their preferred paradigm, their ability to count flashes, \ie visual stimuli, to follow the pace of the interface, and the amount of time they could dedicate for a continuous session while keeping their focus.

\subsection{Signal processing}

After the signal acquisition and the application of temporal filters in relation to the sampling rate (described in \ref{sec:mat}), the P300 embedded in the signal shall be enhanced thanks to some spatial filters to increase its potential detection. Indeed, a usual step for enhancing a particular brain response is to use spatial filters. The P300 is a spatially stationary waveform that has origins different from the background, \ie the current ongoing EEG. In this study, we consider the xDAWN algorithm for creating spatial filters $U$~\cite{rivet09}. The estimation of $U$ is obtained through Rayleigh quotient after two QR decompositions and a singular value decomposition. Every detail can be found in Rivet et al.~\cite{rivet09}. Before the spatial filtering and P300 detection, the signal is first bandpassed filtered (Butterworth filter of order 4) with cut-off frequencies at 1 and 12.5Hz. The signal is then downsampled to 25Hz. The input of the classifier for the P300 detection corresponds to the first four components of the enhanced signal ($N_f=4$). The Bayesian linear discriminant analysis (BLDA) classifier is considered for the detection of the P300 wave~\cite{hoff08}.

\section{Results}

The recorded EEG signal was analyzed offline. Indeed it is possible to analyze data offline as the P300 speller is synchronous, the only difference between the two sessions is the feedback given to the user. This type of analysis is possible as the accuracy is expected to reach more than 90\% online. Moreover, the comparison of the two paradigms is done in the same conditions. The presented results correspond to the cross evaluation between the two sessions of each paradigm. The speller accuracy (in \%) and the ITR (in bits per minute) are presented for each subject in Table~\ref{tab:classif}. The results are given for both paradigms, after 5 and 10 repetitions. The age, gender and experience in using a P300-BCI is mentioned for each subject. For the BCI-P300 experience, 1, 2 and 3 means a BCI naive subject, a subject with experience with few sessions and an experienced user, respectively. The best average accuracy is obtained with XP300 after 10 repetitions with 91.75\%. With CP300, the user needs about 8s to spell one character with 5 repetitions, \ie about 7.5cpm. With XP300, the user would need about 9.3s to spell one character with 5 repetitions, \ie about 6.4cpm. In spite of the lower speed of XP300, the ITR of XP300 is higher than CP300, due to a higher accuracy.

\small

\begin{table*}[htbp]
\begin{center}
\small
\caption{Accuracy (in \%) and ITR (in bpm) for the two paradigms.}
\label{tab:classif}
\begin{tabular}{|cccc|cccc|cccc|}
\hline
        &     &         &        &\multicolumn{4}{|c|}{CP300 speller} & \multicolumn{4}{|c|}{XP300 speller} \\
        &     &         &        & \multicolumn{2}{|c|}{5rep} & \multicolumn{2}{|c|}{10rep} & \multicolumn{2}{|c|}{5rep} & \multicolumn{2}{|c|}{10rep} \\
Subject & Age & Gender & BCI exp & Acc & ITR & Acc & ITR & Acc & ITR & Acc & ITR \\
\hline  
1&	30 & M & 3 & 97.50&	36.64&	100.0&	19.44&	90.00&	26.99&	95.00&	14.91\\
2&	19 & F & 1 & 70.00&	20.68&	80.00&	12.87&	67.50&	16.71&	80.00&	11.03\\
3&	19 & F & 1 & 65.00&	18.35&	87.50&	14.98&	77.50&	20.92&	92.50&	14.18\\
4&	31 & M & 3 & 77.50&	24.41&	95.00&	17.39&	67.50&	16.71&	82.50&	11.61\\
5&	31 & M & 3 & 82.50&	27.09&	95.00&	17.39&	97.50&	31.41&	97.50&	15.70\\
6&	25 & M & 2 & 95.00&	34.79&	100.0&	19.44&	97.50&	31.41&	100.0&	16.66\\
7&	29 & M & 1 & 77.50&	24.41&	90.00&	15.74&	97.50&	31.41&	100.0&	16.66\\
8&	23 & M & 1 & 75.00&	23.13&	90.00&	15.74&	95.00&	29.82&	97.50&	15.70\\
9&	25 & M & 2 & 55.00&	14.05&	65.00&	9.18&		75.00&	19.83&	82.50&	11.61\\
10&	23 & F & 1 & 77.50&	24.41&	90.00&	15.74&	87.50&	25.68&	90.00&	13.50\\
\hline
Mean&	25.5 & - & - &  77.25&	24.80&	89.25&	15.79&	85.25&	25.09&	91.75&	14.16\\
SD  &	4.36 & - & - & 12.06&	6.50&	9.88&	2.91&	11.70&	5.76&	7.25&	2.03\\
\hline
\end{tabular}
\end{center}
\end{table*}	

\normalsize

\begin{figure}
\centering
\subfigure[SA]{\includegraphics[width=.49\columnwidth,trim = 5mm 5mm 20mm 10mm, clip]{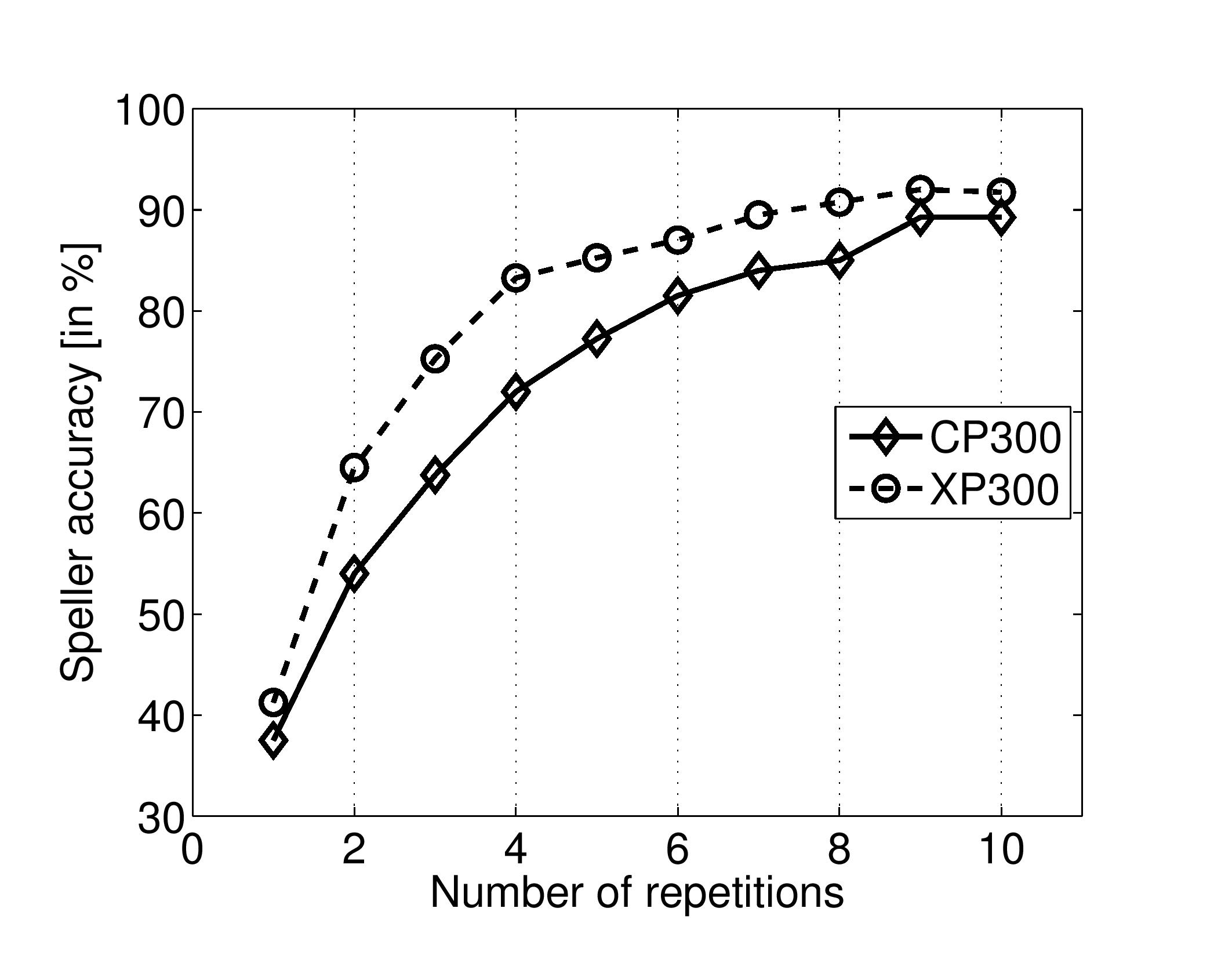}\label{fig:reco}}
\subfigure[ITR]{\includegraphics[width=.49\columnwidth,trim = 5mm 5mm 20mm 10mm, clip]{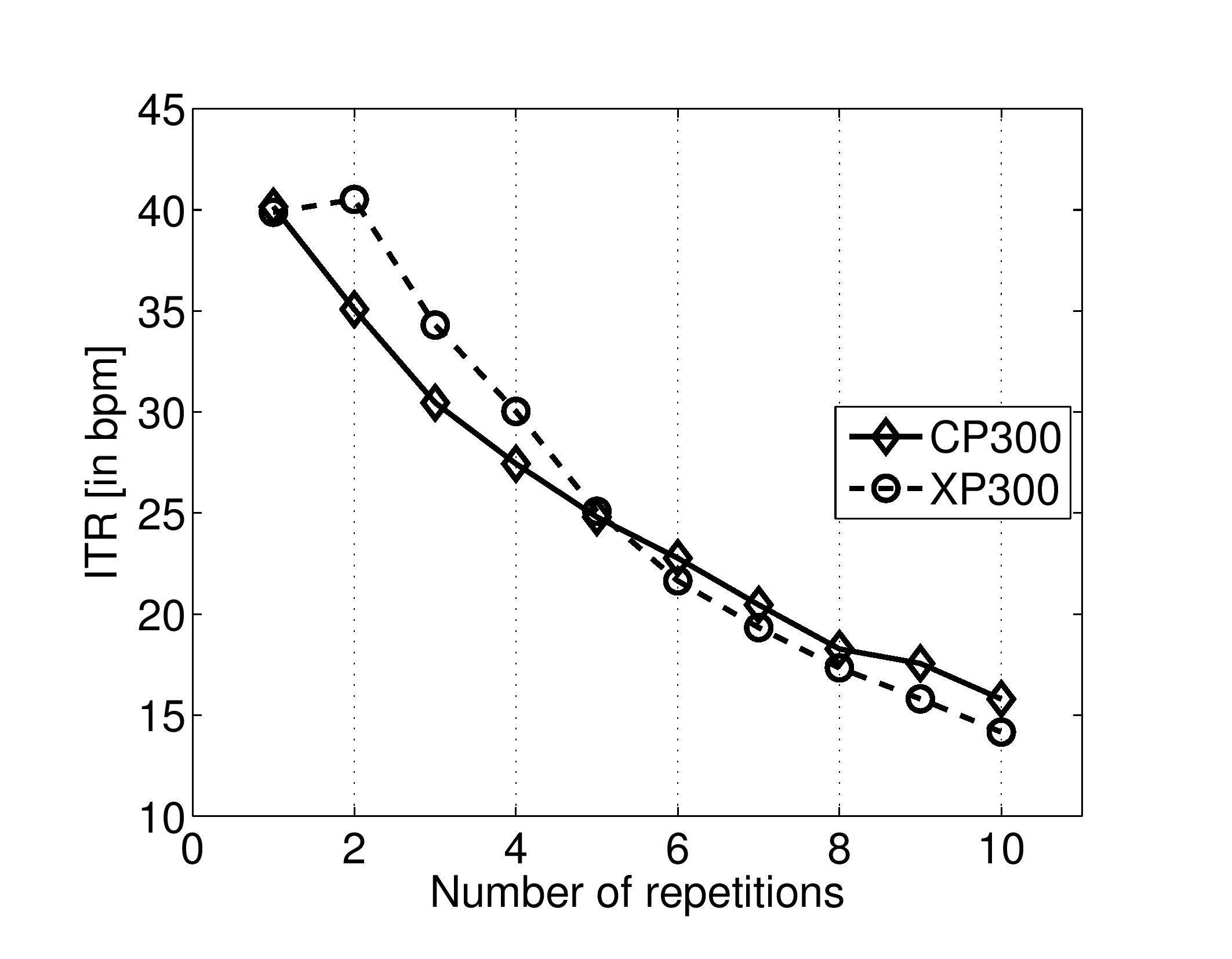}\label{fig:itr}}
\caption{Comparison of the performance achieved by classical (CP300) and proposed (XP300) paradigms.
Fig.~\ref{fig:reco}: average speller accuracy (SA) across repetitions.
Fig.~\ref{fig:itr}: average information transfer rate (ITR) across repetitions.}
\end{figure}

\begin{figure}
\centering
\subfigure[CP300]{\includegraphics[width=.49\columnwidth,trim = 5mm 5mm 5mm 10mm, clip]{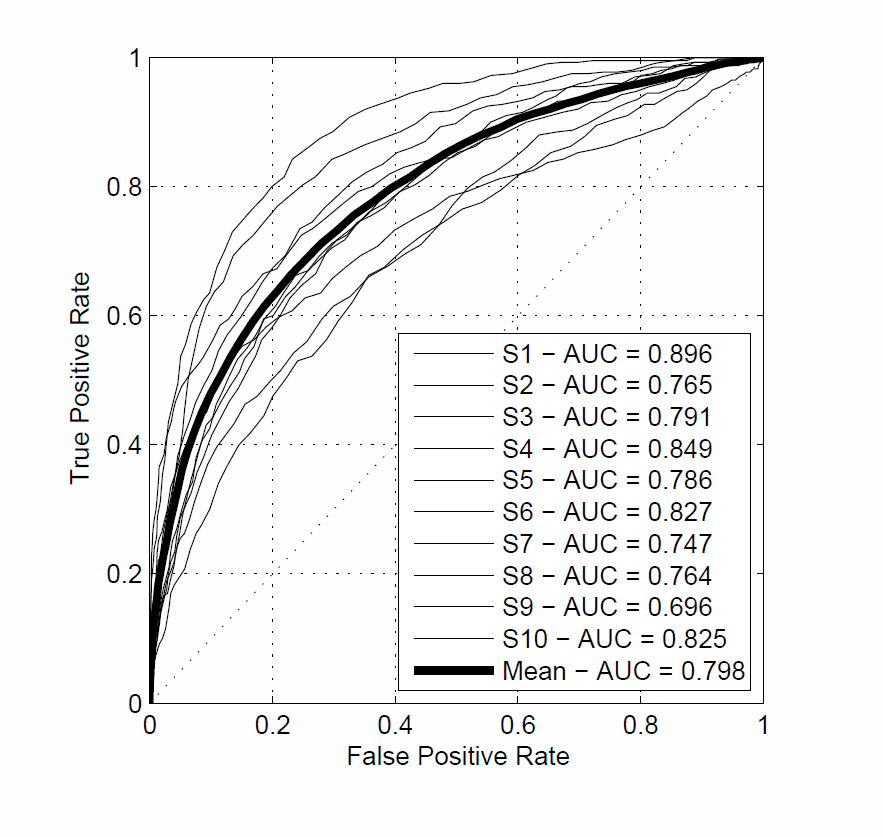}\label{fig:CP300}}
\subfigure[XP300]{\includegraphics[width=.49\columnwidth,trim = 5mm 5mm 5mm 10mm, clip]{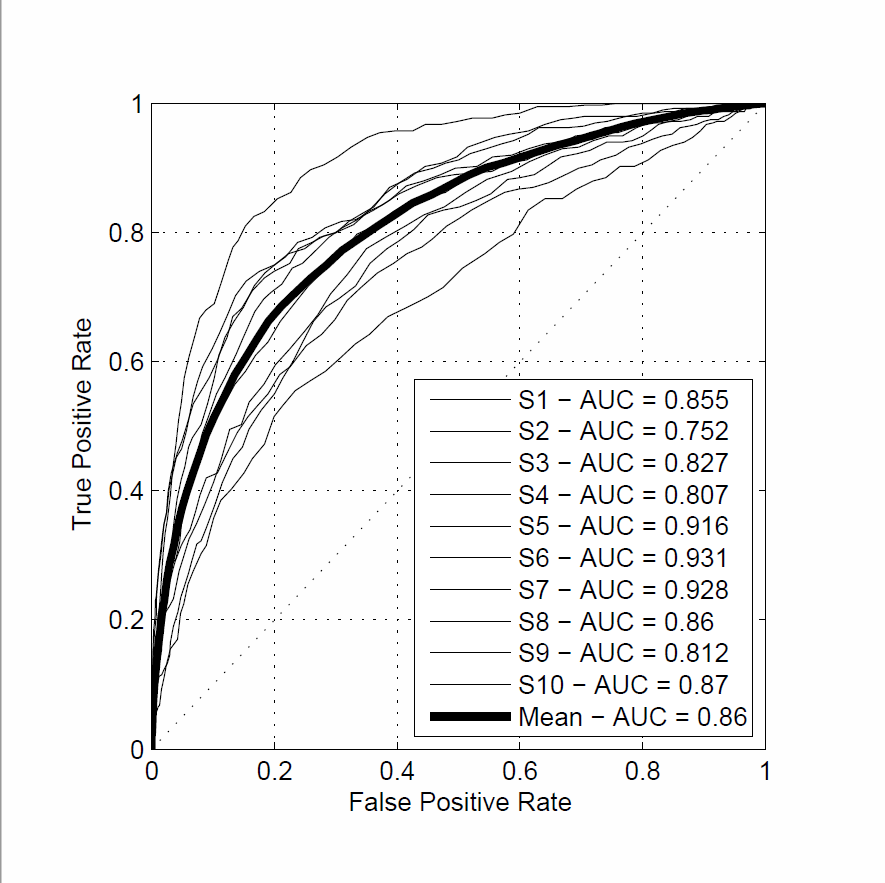}\label{fig:XP300}}
\caption{Comparison of the ROC curves of the P300 detection for CP300 (Fig.~\ref{fig:CP300}) and XP300 (Fig.~\ref{fig:XP300}).}
\label{fig:roc}
\end{figure}

The average speller accuracy and ITR across repetitions are depicted in Fig.~\ref{fig:reco} and~\ref{fig:itr}, respectively. As the visual stimuli are fast and the recognition with one repetition is close to 50\%. The best trade-off between speed and accuracy is at 4 or 5 repetitions. As a short pause between each repetition is considered with XP300, the ITR with XP300 gets inferior than with CP300 as the accuracy reaches more than 90\% in both cases. A pairwise t-test comparison between CP300 and XP300 shows that the mean accuracy across repetitions is higher with XP300 ($t_9=-6.18$, $sd=3.42$, $p=8.08e-05$). Across subjects, a pairwise t-test comparison reveals that XP300 has a higher accuracy when we consider 3, 4 or 5 repetitions; their corresponding p-values are 0.0209, 0.0073 and 0.0285. However, a paired t-test failed to reveal significant differences in the ITR across subjects. Single trial detection has also been evaluated. Figure~\ref{fig:roc} presents the ROC (Receiver Operating Characteristic) curve for every subject and the mean curve across subjects for CP300 and XP300. A pairwise t-test comparison indicates that the AUC (Area Under the Curve) is higher with XP300 ($t_9=-2.79$, $sd=0.07$, $p=0.019$).

\begin{figure}
\centering
\begin{tabular}{lcc}
$FC_Z$ & \includegraphics[width=0.4\columnwidth,trim = 1mm 5mm 10mm 5mm, clip]{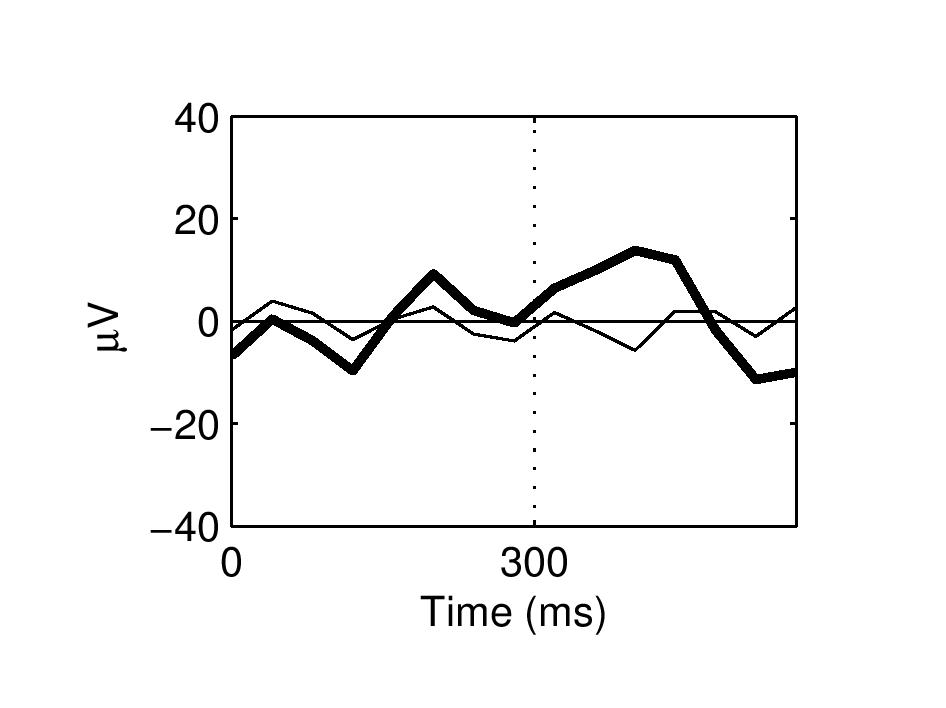} & \includegraphics[width=0.4\columnwidth,trim = 1mm 5mm 10mm 5mm, clip]{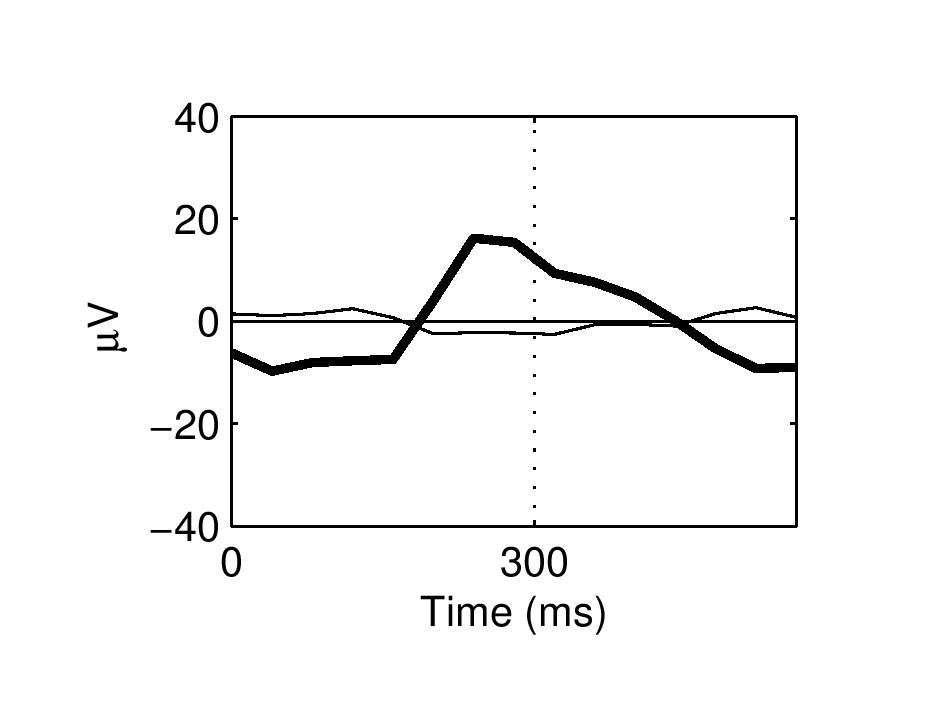} \\
$P_Z$ & \includegraphics[width=0.4\columnwidth,trim = 1mm 5mm 10mm 5mm, clip]{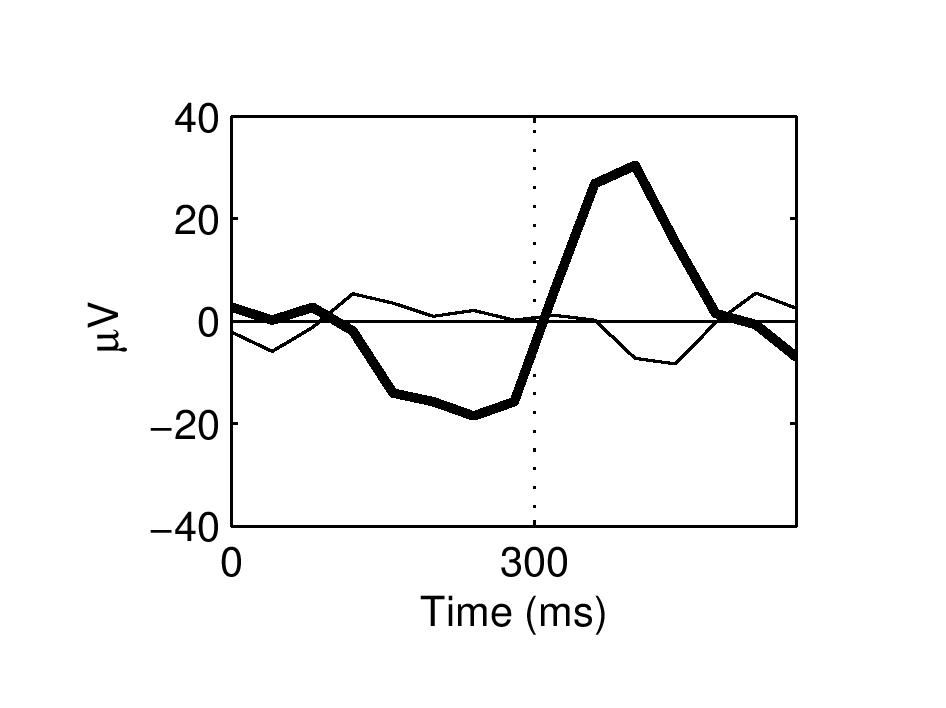} & \includegraphics[width=0.4\columnwidth,trim = 1mm 5mm 10mm 5mm, clip]{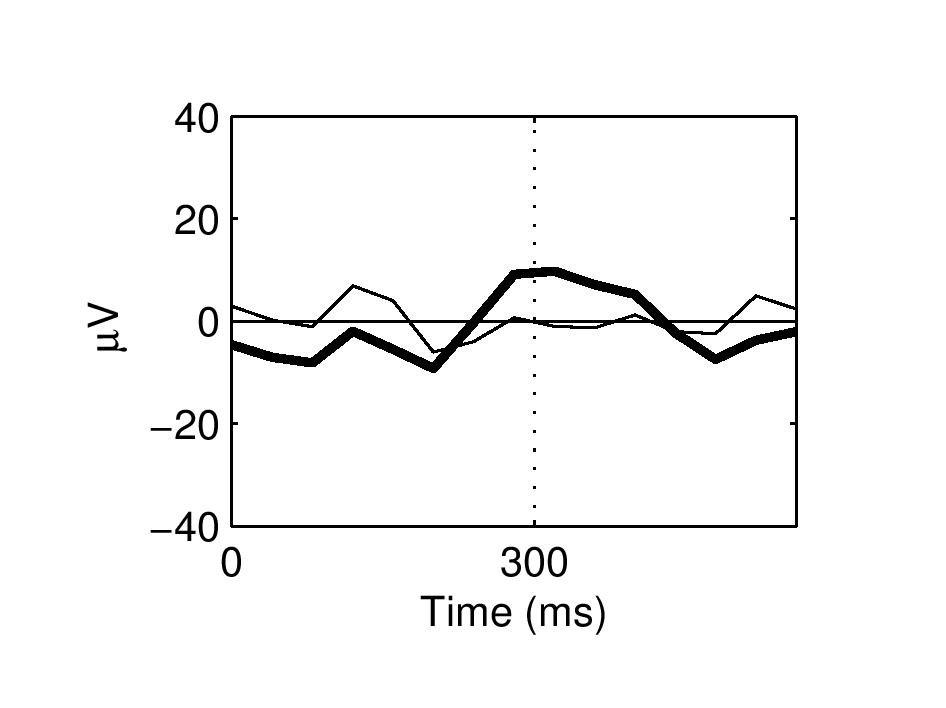} \\
$P_3$ & \includegraphics[width=0.4\columnwidth,trim = 1mm 5mm 10mm 5mm, clip]{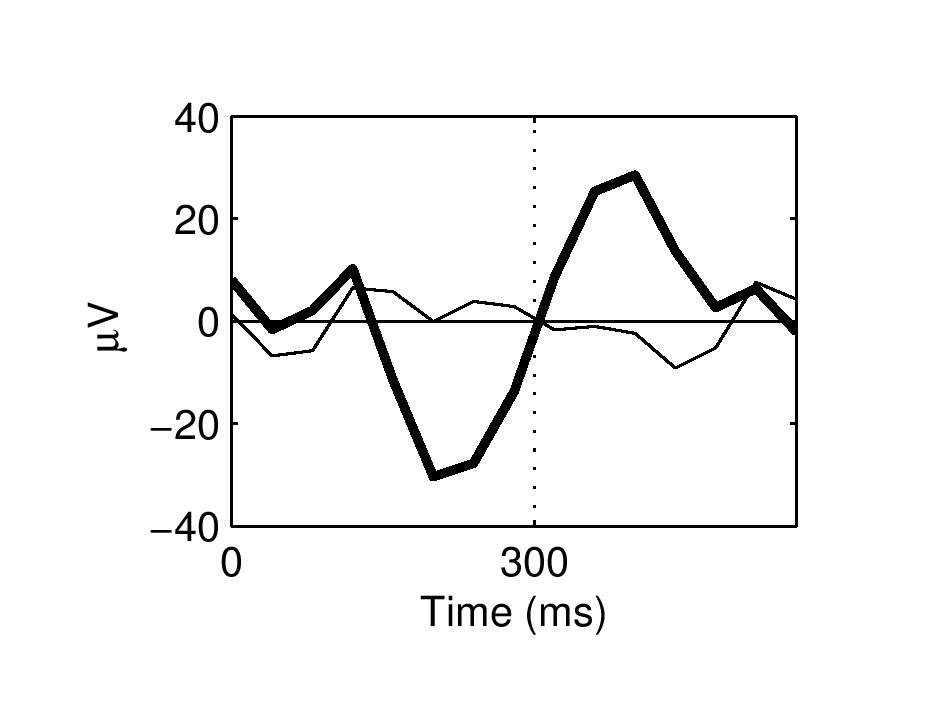} & \includegraphics[width=0.4\columnwidth,trim = 1mm 5mm 10mm 5mm, clip]{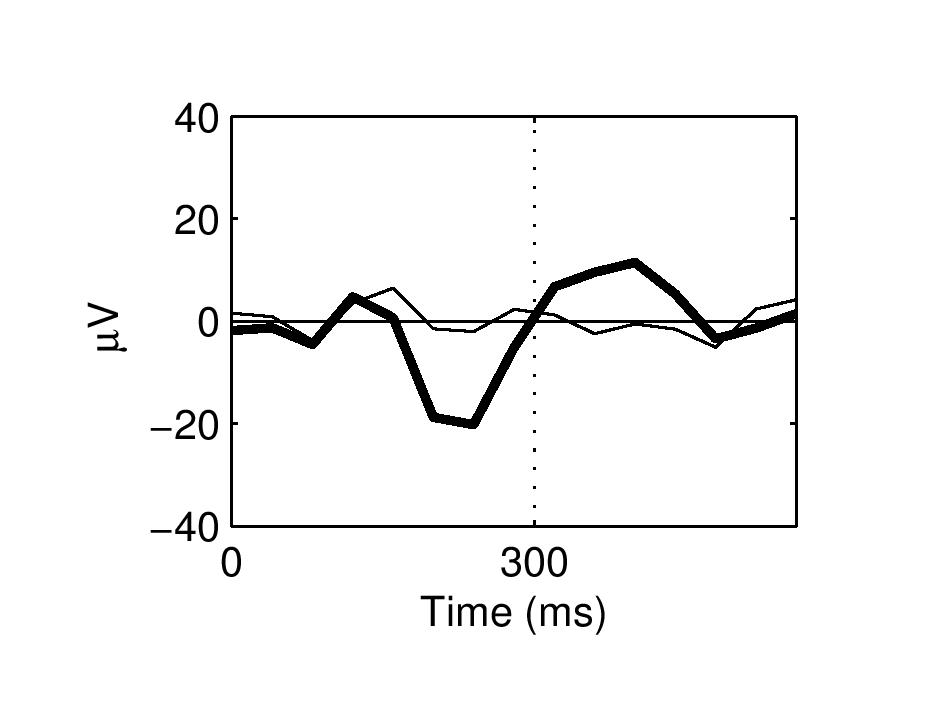} \\
$P_4$ & \includegraphics[width=0.4\columnwidth,trim = 1mm 5mm 10mm 5mm, clip]{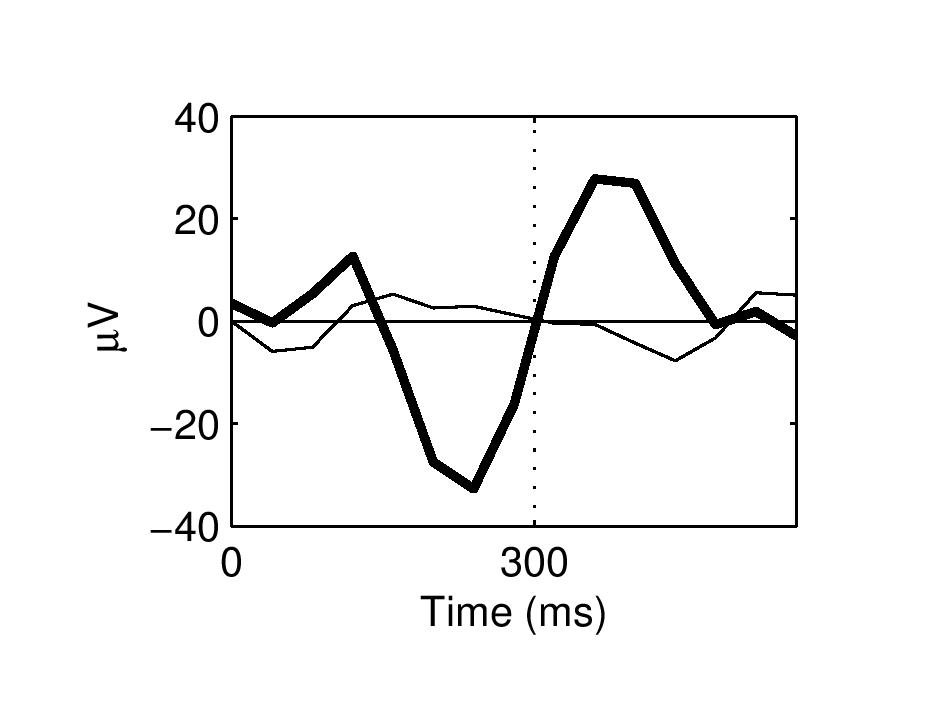} & \includegraphics[width=0.4\columnwidth,trim = 1mm 5mm 10mm 5mm, clip]{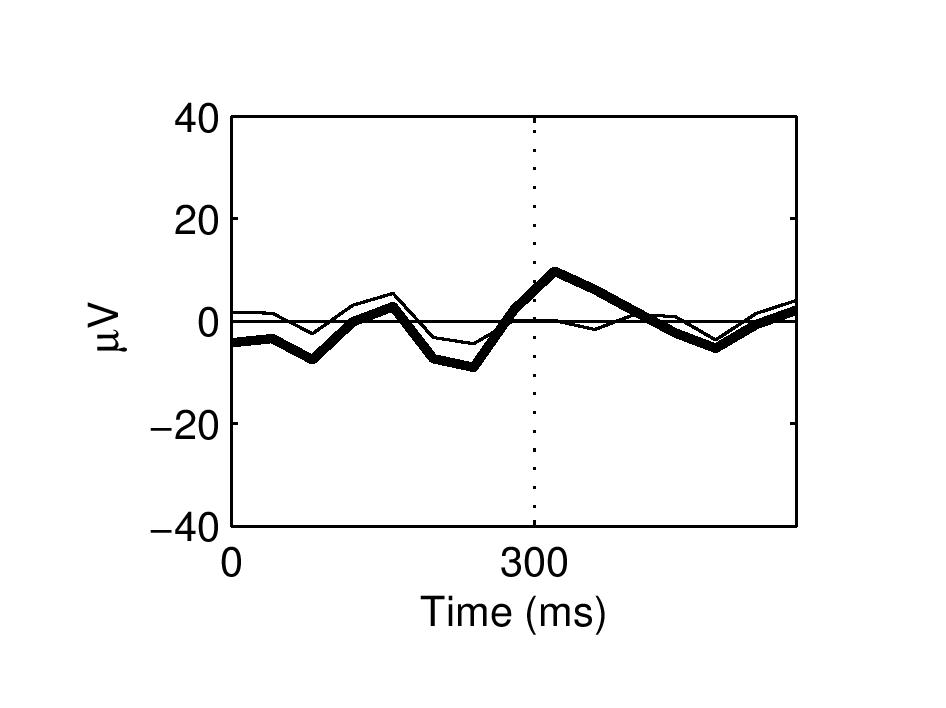} 
\end{tabular}
\caption{Averaged ERP on the target and non-target across the different electrodes, for CP300 (left) and XP300 (right), starting after a visual stimulus and during 0.6s. The bold and thin lines represent the ERP on the target and non target, respectively.}
\label{fig:waveA}
\end{figure}

\begin{figure}
\centering
\begin{tabular}{lcc}
$P_7$ & \includegraphics[width=0.4\columnwidth,trim = 1mm 5mm 10mm 5mm, clip]{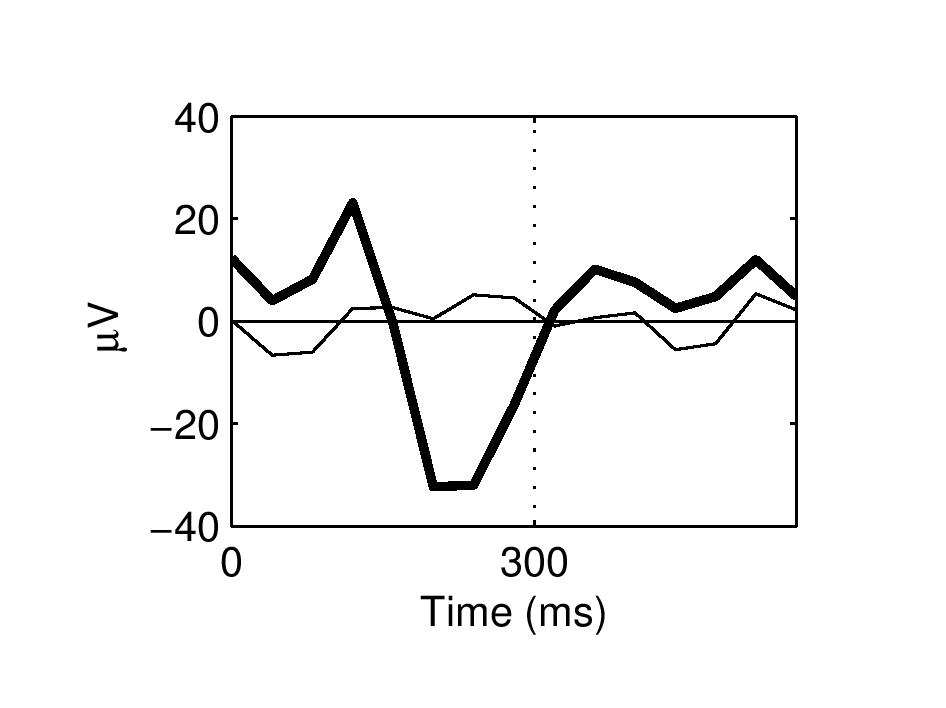} & \includegraphics[width=0.4\columnwidth,trim = 1mm 5mm 10mm 5mm, clip]{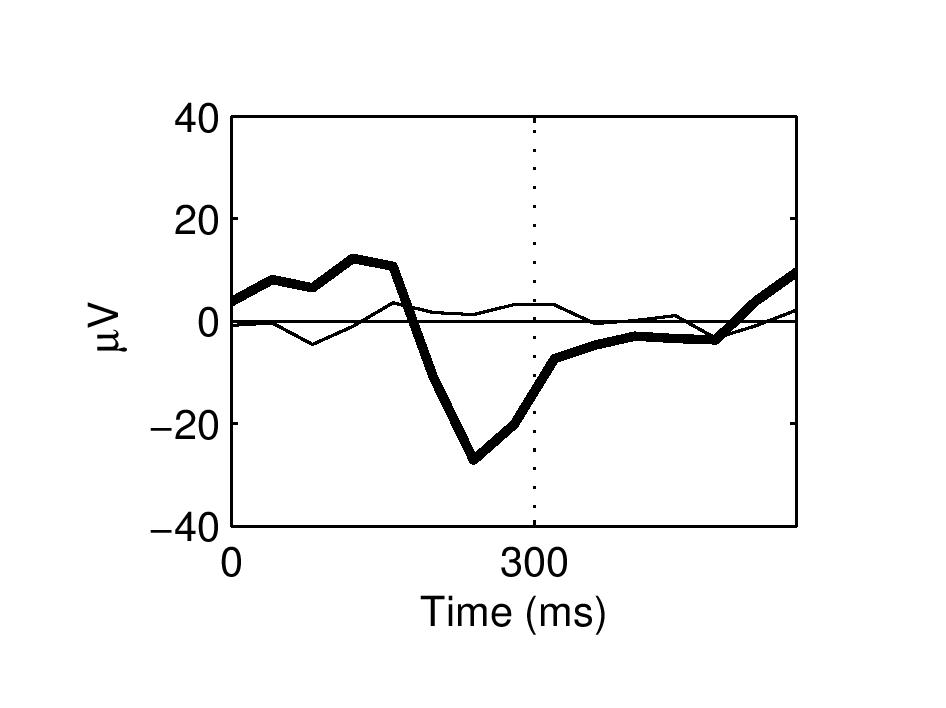} \\
$P_8$ & \includegraphics[width=0.4\columnwidth,trim = 1mm 5mm 10mm 5mm, clip]{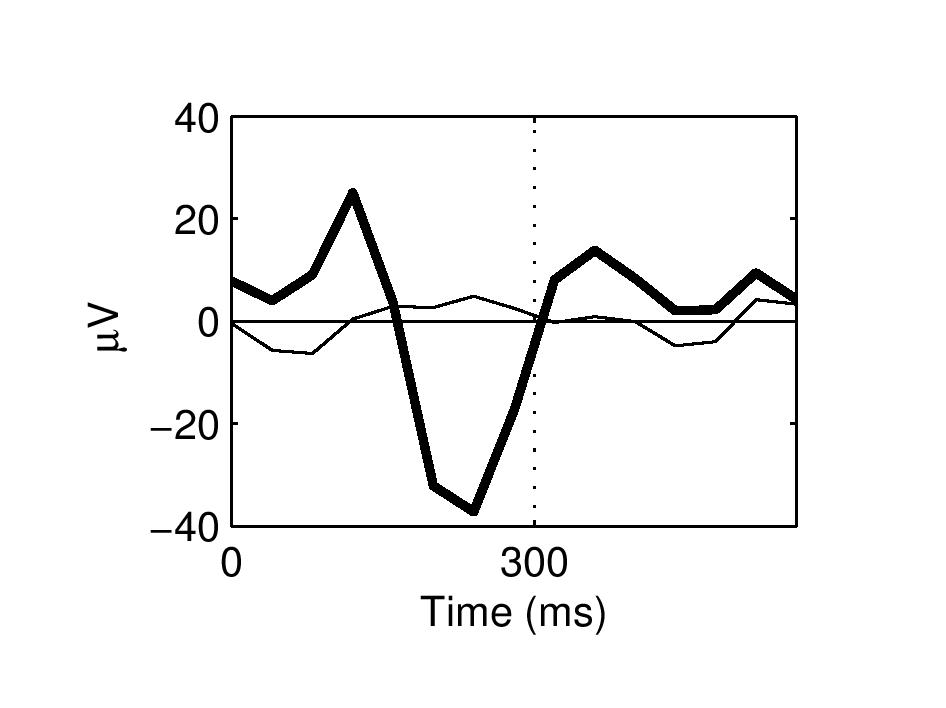} & \includegraphics[width=0.4\columnwidth,trim = 1mm 5mm 10mm 5mm, clip]{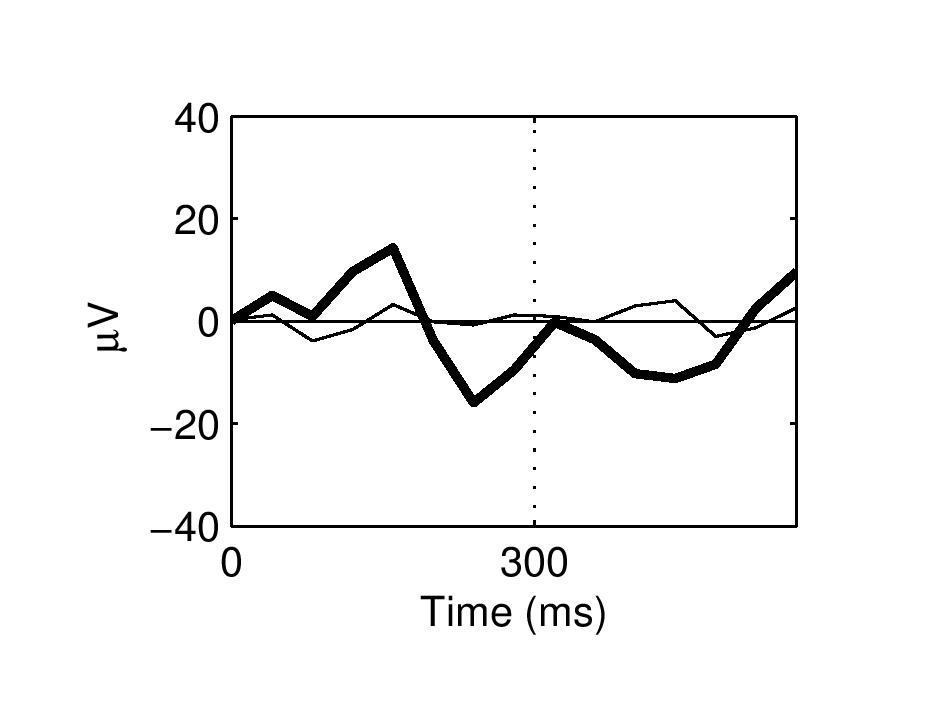} \\
$O_1$ & \includegraphics[width=0.4\columnwidth,trim = 1mm 5mm 10mm 5mm, clip]{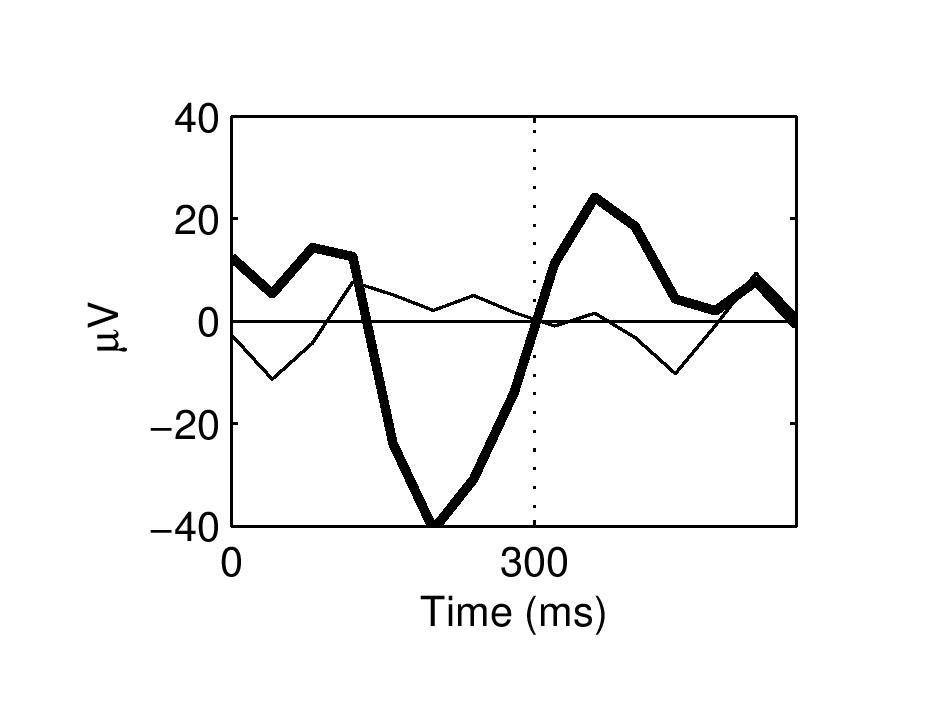} & \includegraphics[width=0.4\columnwidth,trim = 1mm 5mm 10mm 5mm, clip]{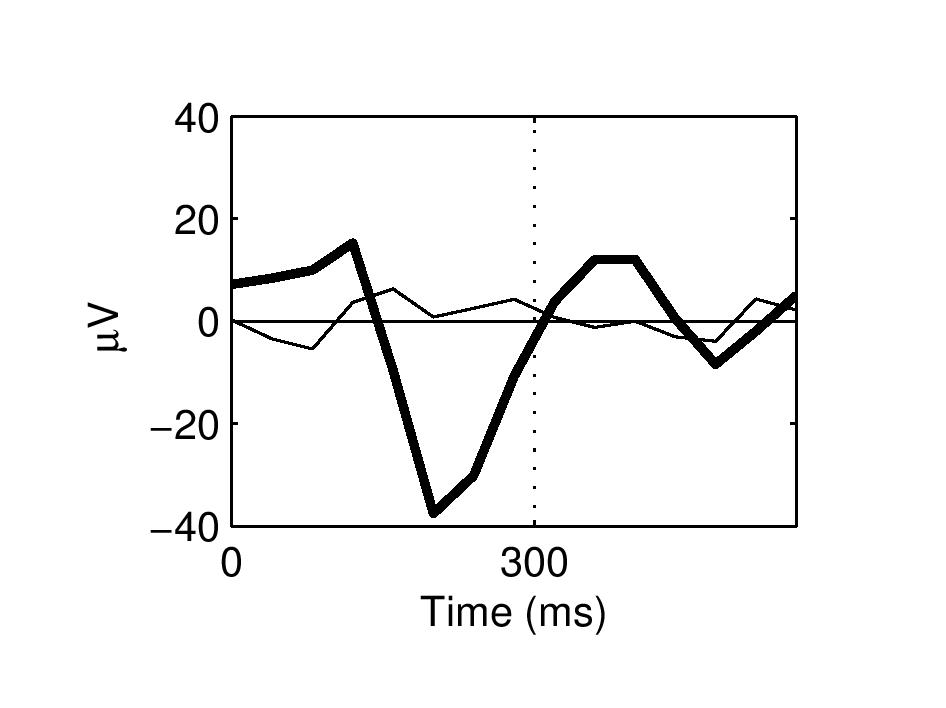} \\
$O_2$ & \includegraphics[width=0.4\columnwidth,trim = 1mm 5mm 10mm 5mm, clip]{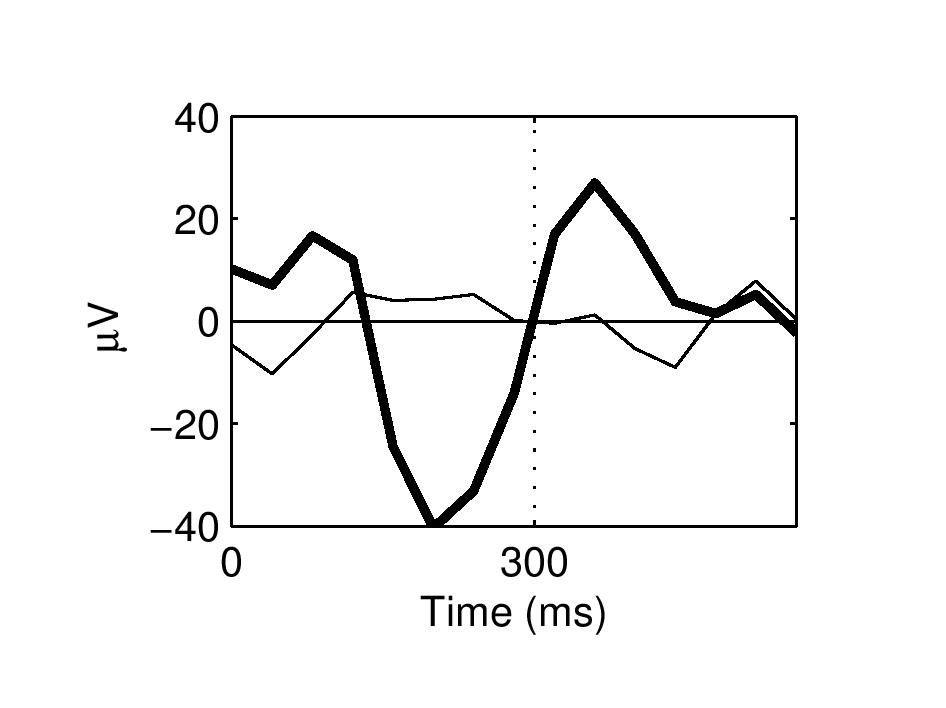} & \includegraphics[width=0.4\columnwidth,trim = 1mm 5mm 10mm 5mm, clip]{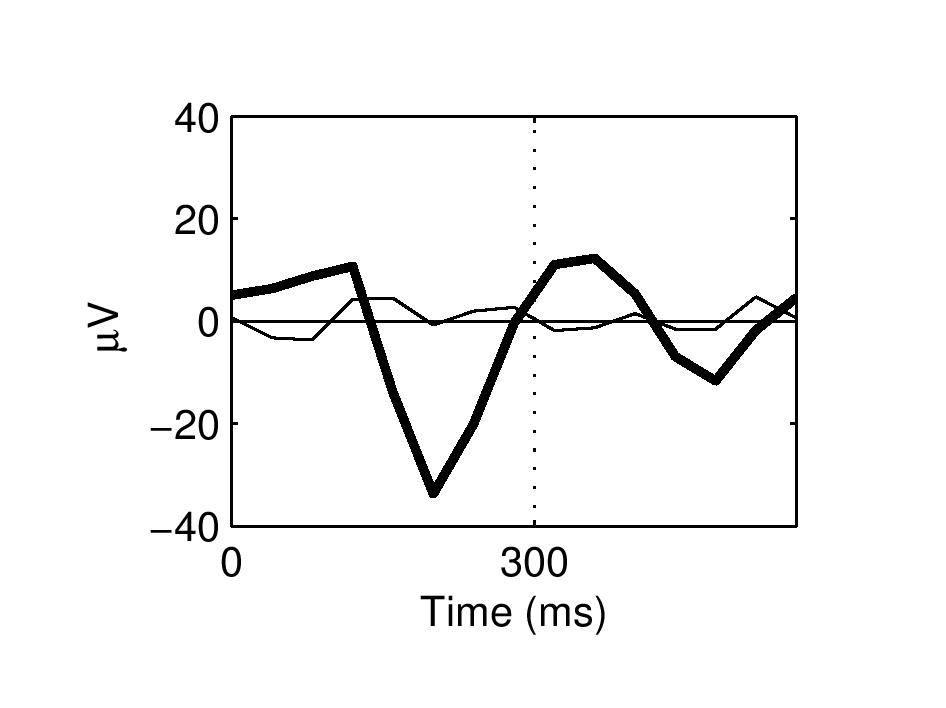}
\end{tabular}
\caption{Averaged ERP on the target and non-target across the different electrodes, for CP300 (left) and XP300 (right), starting after a visual stimulus and during 0.6s. The bold and thin lines represent the ERP on the target and non target, respectively.}
\label{fig:waveB}
\end{figure}

The averaged ERP, after temporal filtering, are presented in Fig.~\ref{fig:waveA} and~\ref{fig:waveB} (the thin lines represent the non target ERP). These figures display the importance of the visual components. The starting point corresponds to the beginning of a visual stimulus. Several differences can be observed between CP300 and XP300. In both paradigms, the N200 component is high on $O_1$ and $O_2$. The N200 component is less important on $P_4$ and $P_7$ with XP300 than CP300. On $FC_Z$, the P300 is more visible with XP300 than with CP300. This difference can be explained by the different visual stimuli that allow counting flashes.

In addition to the raw performance of the speller, an important question is to know which paradigm the users would choose. According to the questionnaire results and the subjects comments, XP300 was preferred by every subject. In the comments, the type of fatigue differs in relation to the paradigm. With CP300, it is harder to pay attention to the flashes. Thus, subjects feel the need to be more focused with CP300. With XP300, every subject feels more comfortable during the experiment. However, some subjects acknowledge that the feedback for counting flashes could be an inconvenient: when the user feels comfortable with the interface, their mind may wander. In spite of these remarks, XP300 was chosen by every subject as the best choice for potential long sessions. Each flash has a meaning and it is possible to shift directly to the next character once a character is spelt. While the speller accuracy is only better with XP300 with a low number of repetitions due to the efficiency of the P300 detection, the subjective evaluation of XP300 clearly shows the user preference of XP300.

\section{Discussion and conclusion}

The proposed method is a good alternative to the classical P300 paradigm. Indeed, it increases the oddball paradigm aspect by considering pseudo-random subsets of symbols for each flash instead of rows and columns. The odd aspect of the oddball paradigm has been limited to the order of the flash (odd in time), but here it is formally spatially odd, without increasing the number of visual stimuli. As the visual stimuli with XP300 are presented in an odder way than with CP300, it is possible to separate the blocks of flashes and reduce the probability of consecutive flashes on the target. In addition, the feedback for counting the visual stimuli is a great help for subjects by increasing the meaning of the visual stimuli. These encouraging results would probably benefit the current target of P300-BCI users, \eg locked-in patients. Many graphical improvements could benefit the original P300 speller paradigm. These improvements should be oriented toward increasing the interface usability. Like for signal processing methods, some parameters can be set for all subjects, \eg bandpass filtering, and others shall be tuned in relation to the user, \eg spatial filters. The proposed paradigm proposed solutions to hypotheses that were already validated in the literature: the confusions around the target due to the RC paradigm, the attentional blink when the target to target interval is too short, and the importance of the stimuli meaning. Although the proposed solution provided better results across subjects, some parameters may be set in relation to a particular individual. 

The characteristics of the paradigm are sources for better understanding the underlying neural processes in the P300 speller. However, the observations of the ERPs suggest a major role for the visual evoked potentials. The modifications of the visual stimulus shall be defined as parameters that affect the P300 amplitude: subjective probability, stimulus meaning, and information transmission~\cite{joh86}. The visual stimulus in the P300 paradigm has no meaning for a person who wants to spell a character. This is due to the number of repetitions that are done for the selection of a character. Indeed, a single visual stimulus does not imply the selection of a character. The relevance of the stimulus to the task should be increased. The P300 amplitude is lower when some stimulus information is lost, like when the stimulus is harder to discriminate or perceive. P3b requires attention, and increasing the difficulty of maintaining attention will correspondingly decrease P3b amplitude. The relevance of the stimulus to the task, and the amount of information a stimulus transmits are all variables that will determine P3b amplitude~\cite{joh86}. In XP300, the stimulus is fully integrated as a part of the character selection: the little dots around the character shall be filled for its selection. The little dots work here as a feedback for the main stimulus (the change of color and size of the matrix cell). This could suggest that each individual stimulus should have a meaning and should be translated into action. As not only perceptual processes are involved, the user should be able to assign a meaning to the visual stimuli, which are related to the application. Other feedbacks, like neurofeedback in relation to single trial detection, could probably enhance the meaning of each visual stimulus. 

Finally, although the P300 speller can still be improved through methods that increase the reliability over time and across subjects, and the ITR, future works will directly be addressed at the application level \eg by considering word completion, prediction, and knowledge of the vocabulary.  


\small{
\bibliographystyle{unsrt}
\bibliography{bibliox}
}

\end{document}